\documentclass[12pt]{iopart}
\usepackage[english]{babel}
\usepackage[T1]{fontenc}
\usepackage{microtype}
\usepackage{graphicx}
\usepackage[usenames,dvipsnames]{color}
\usepackage{hyperref}
\usepackage{float,amssymb}

\newcommand{\omlas}{\omega}
\newcommand{\wcmq}{\mathrm{W}/\mathrm{cm}^2}
\newcommand{\nesc}{\mathrm{N}_\mathrm{esc}}
\newcommand{\tpulse}{{T_\mathrm{pulse}}}

\newcommand{\ekin}{E}

\newcommand{\veck}{\mathbf{k}}
\newcommand{\vece}{\mathbf{e}}
\newcommand{\vecr}{\mathbf{r}}
\newcommand{\veca}{\mathbf{A}}

\newcommand{\mbraket}[3]{\langle #1|#2|#3\rangle}

\newcommand{\rbox}{r_\mathrm{box}}
\newcommand{\tsurff}{t--SURFF}

\usepackage[normalem]{ulem}

\begin{document}

\title[Modeling of electron photoemission in 
nanostructures with TDDFT]{Efficient and accurate modeling of electron photoemission in 
       nanostructures with TDDFT}
  
\author{Philipp~Wopperer$^1$, Umberto~De Giovannini$^1$, and Angel~Rubio$^{1, 2}$}

\address{$^1$ Nano-Bio Spectroscopy Group and ETSF, Universidad del Pa\'is Vasco, 
         CFM CSIC-UPV/EHU, 20018 San Sebasti\'an, Spain.}
\address{$^2$ Max Planck Institute for the Structure and Dynamics of Matter and Center for 
         Free-Electron Laser Science, Luruper Chaussee 149, 22761 Hamburg, Germany.}
\ead{philipp.wopperer@ehu.es}
% \vspace{10pt}
% \begin{indented}
% \item[]August 2016
% \end{indented}

\begin{abstract}
  We review different computational methods for the calculation of photoelectron spectra and 
  angular distributions of atoms and molecules when excited by laser pulses using 
  time-dependent density-functional theory (TDDFT) that are suitable for the description
  of electron emission in compact spatial regions. We derive and extend the time-dependent 
  surface-flux method introduced in Reference~\cite{Tao12} within a TDDFT formalism and 
  compare its performance to other existing methods.
  We illustrate the performance of the new method by simulating strong-field ionization 
  of C$_{60}$ fullerene and
  discuss final state effects in the orbital reconstruction of planar organic molecules.
\end{abstract}

% \maketitle

\vspace{2pc}

\section{Introduction}
When physical systems such as atoms, molecules, clusters, or nano-objects 
are exposed to an appropriately tuned radiation field there is a non-negligible 
chance of ionization. Strictly speaking, electron photoemission takes place whenever 
the exciting 
field is capable to induce a bound-to-continuum transition and results in electrons 
escaping with a given kinetic energy and from a given direction.
Depending on the characteristics of the radiation, the target system, 
time and energy scales, there exist many mechanisms that contribute to emit electrons and 
correspondingly many ways to categorize them: for example, 
sequential (Fano resonances, Auger decay, autoionization/thermionic emission) 
and non-sequential processes, single and double ionization, 
single- and multiphoton ionization, above-threshold ionization, etc.
Associated with each of these mechanisms there is as much spectroscopic information
to be extracted.
In particular, knowledge about the ionization dynamics and on the parent system can be 
gained from differential observables of the electron yield. For instance, 
the electron photoemission probability as a function of the kinetic energy -- the photoelectron 
spectrum (PES) -- can give insight into electronic energy levels, while the 
photoelectron angular distribution (PAD)
carries spatial information on ionic positions or directly on the electronic configurations.

If we focus on the ionization mechanism as a function of the external field parameters, we 
can distinguish among different regimes.
A common classification of the predominant ionization regimes
is according to the 
Keldysh parameter~\cite{Kel65}, originally defined for the hydrogen atom. 
This parameter is given by $\gamma = \sqrt{E_\mathrm{IP}/(2U_p)}$ with 
$E_\mathrm{IP}$ being the ionization potential of the system, 
$U_p = I/(4\omlas^2)$ the ponderomotive energy~(i.e., the averaged quiver energy in atomic units), 
  $I$ the intensity of the irradiating light, and 
  $\omlas$ the frequency.
For $\omega\ll E_\mathrm{IP}$, 
one distinguishes two ionization regimes according to the Keldysh parameter:
the perturbative or nonlinear multiphoton ionization regime~(MPI)~\cite{Man91} which is associated 
with $\gamma > 1$ and 
the strong-field ionization regime with $\gamma < 1$.

In the multiphoton regime, the laser action results in a vertical 
excitation of a bound electron into the continuum by absorption of several photons.
Nevertheless, weak and moderate 
lasers can promote electrons also far into the continuum by absorption 
of multiple photons above the ionization threshold.
This non-perturbative process is called above-threshold ionization 
(ATI)~\cite{Ago79} and yields to
spectra which decay exponentially in energy and that are characterized by a series of peaks
separated by $\omega$.

In the past decade, research in the photoelectron spectroscopy of finite systems
has focused predominantly on the strong-field regime which is characterized 
by the onset of optical field ionization. At sufficiently high field strengths, 
the barrier of the binding Coulomb potential is suppressed which results in a 
tunneling current that follows adiabatically the variation of the laser field~\cite{Bra00}.
Above-threshold ionization still prevails in this regime, however, it is 
more suitably explained as interferences between coherent photoelectron 
wavepackets emitted at different 
times within the laser cycle. 

Photoelectron spectra and angular distributions in the strong-field regime are particularly rich in 
information on ionization dynamics since field-driven rescattering is involved. 
This process was described by a classical three-step model, also called 
simpleman's model~\cite{Cor93, Kul93}
and was later extended to a quantum description (strong-field approximation)~\cite{Lew95} 
using the 
Keldysh--Faisal--Reiss approximation~\cite{Kel65, Fai73, Rei80}. According to the three-step model
the electron is first released by tunnel ionization. The tunneled electron is then 
accelerated in the laser field where it acquires a kinetic energy. 
Depending on the release time, electrons either leave the parent ion directly 
or are driven back towards the parent ion once the laser field changes its sign~\cite{Kra09}. 
Applications of the field-driven rescattering process are manifold.
For instance, rescattering coherent electron wavepackets can be used to 
self-interrogate the parent molecular structure. This phenomenon is exploited for 
laser-induced electron diffraction~(LIED)~\cite{Mec08, Bla12} to image molecular structures. 
Rescattering and direct photoelectron wavepackets can also interfere coherently which allows 
to study holographic pattern in photoemission spectra and angular distributions~\cite{Huismans2011}.
Nevertheless, reading recollision induced diffraction images can become a complex task 
as several processes compete on similar time and energy scales~\cite{Spa04}.

Concerning materials and investigated species in photoemission experiments, research
has extended to larger and denser systems in the 
previous years. Quite recently,
signatures of strong-field physics were found by experiments in metallic 
surfaces~\cite{Rac11} and
nanostructures~\cite{Dom13a}, clusters~\cite{Fen07}, and dielectric nanospheres~\cite{Zhe11}.
For instance, the exploration of photoionization processes in metal nanotips is currently a 
strongly evolving field of research. The combination of femtosecond laser pulses and a sharp 
metal tip is considered as a laser-driven ultrafast electron emitter on the nanometer scale 
with prospective applications as electron source in electron microscopy, electron diffraction 
and for free electron lasers, as an extremely sensitive carrier-envelope phase sensor, or as 
generators of high-harmonic radiation~\cite{Pig13a,Kru11,Her12a, Bio13}.

Calculation of PADs from the perturbative to the strong-field regime, 
and accurate modeling of photoemission experiments on a broad range of materials at the same time, 
can only be achieved by a comprehensive approach.
In general, the interaction between electrons in an atom or a molecule and a laser field 
is difficult to treat theoretically, and several approximations are usually 
employed. For one-electron systems, PES and PAD can be calculated exactly by
directly solving the time-dependent Schr\"odinger equation~(TDSE). 
The most straightforward way is 
by projecting the wavefunction obtained 
from the TDSE at the end of 
the pulse onto continuum states~\cite{Bac01}. 
Another approach where the calculation of the continuum eigenstates is 
avoided, is the resolvent technique~\cite{Cat12}. Both methods need to propagate the wavefunction 
until the end of the pulse in a large space domain in order to obtain the correct 
distribution of the ejected electrons.
For simple cases this problem can be overcome by the use of 
spherical coordinates. Also geometrical splitting techniques~\cite{Che98, Ton07, He15}
turn out to be very useful to reduce computational cost.

For more than two electrons, the exact solution of the 
TDSE in three dimensions is unfeasible and basically all ab-initio 
calculations for multielectron systems are performed under the single-active 
electron (SAE) approximation. 
In the SAE only one electron interacts with the external field while the other electrons 
are frozen. This approximation was successfully employed in several photoemission studies for atoms and molecules 
in strong laser fields~\cite{Che06, Awa08, Pet10}. 
Besides the TDSE, Floquet theory~\cite{Mad97, Chu04}, 
the strong-field approximation~\cite{MB00, Dre14}
and semi-classical methods~\cite{Yud01, Ton02, Dim04, Eck08} based on ionization rates~\cite{ADK86}
are used in the strong-field regime.
For weak lasers, plane wave methods~\cite{Pus09}, the independent 
atomic center approximation~\cite{Gro78} 
and (multiphoton) perturbation theory~\cite{Tof12, Sei02, Fai87} are usually employed.
However, such approaches reproduce dynamics only qualitatively, and 
their failure to describe 
multielectron (correlation) effects and their often oversimplified 
assumptions for the continuum state call for better schemes.

The inclusion of exchange-correlation effects for a system of many interacting electrons 
can be achieved within time-dependent density-functional theory (TDDFT)~\cite{RG84, Fun12}.
Computations of electronic excitations for systems with up to a few hundred atoms 
are currently most widely carried out employing this method. TDDFT offers a 
reasonable trade-off between accuracy and computational cost, where other, more accurate, 
methods~\cite{For92, Zan04} would not be feasible. 
In spite of 
transferring all the many-body problems into an unknown exchange-correlation functional, 
the calculation of PES and PAD in TDDFT is not straightforward. While the total ionization yield 
can be calculated directly, differential quantities cannot be expressed 
in terms of the electron density. 
Methods based on TDDFT, therefore, assume that PES and PAD can be directly obtained 
from the time-dependent Kohn-Sham orbitals. 
Nevertheless, a close correspondence between spectroscopic data and Kohn-Sham orbitals exists 
when using self-interaction-free exchange and correlation functionals~\cite{Dauth2011}.

For TDDFT there are only two methods to compute PES and PAD that can be formulated in 
finite volumes and that do not require to explicitly calculate continuum states~\cite{Larsen:2015kc}: the sampling point method~(SPM)~\cite{Poh00, Di12b} and the 
mask method~(MM)~\cite{deGio12}. Both of them were extensively and successfully 
used for the calculation of photoelectron spectra of 
atoms, molecules, and clusters~\cite{Wop14a}, model systems for 
nano-tips~\cite{Wac12a}, and for various experimental 
setups from time-resolved (pump-probe) spectroscopy~\cite{DeGio13, CrU14b} to 
strong-field ionization of atoms exposed the x-rays~\cite{CrU14a}. 
However, both methods present limitations in practical applications.
For instance, MM becomes demanding to converge for low kinetic energies ($E\lesssim 1$~eV). 
This is because MM requires Fourier transforms of the wavefunctions.
Thus, small energy steps and consequently small momenta are associated with large spatial dimension 
that become increasingly large as we decrease the step. 
Furthermore, the use of Fourier transforms prevents efficient parallelization in spatial domains which 
in turns limits the size of the largest simulation box to a single computational node memory. 
On the other end, SPM is less limited from the computational stand point, but is unreliable especially in the strong-field regime. 
This is due to the strong assumptions that it needs which are difficult to assess and in turn require comparatively large simulations boxes to appropriately converge. 

A promising alternative method was proposed in Ref.~\cite{Tao12, Scr12}, 
and a preliminary version of the same method in Ref.~\cite{Cai05} -- the time-dependent 
surface flux method (\tsurff). 
\tsurff\ has been so far employed only for few-electron systems either with 
TDSE~\cite{Tao12, Scr12,Morales:2016va} or in combination with multiconfigurational time-dependent 
Hartree-Fock~\cite{Cai05}.
In this paper, we extend this method for the first time to TDDFT. In Sec.~\ref{sec:theory},  
we present the theory alongside the sampling point and mask methods to illustrate differences 
and common traits and proceed with a real world comparison on a characteristic set of examples 
in Sec.~\ref{sec:examples}.
For the sake of simplicity, we restrict ourselves here and following 
to spin-unpolarized many-electron systems. Nevertheless, all 
expressions and calculations can be trivially extended to include spin polarization.

Atomic units will be used throughout ($m_e = e = \hbar =1$) unless otherwise
indicated.

\section{Theory}\label{sec:theory} 
\subsection{Space partitioning and momentum distribution}\label{sec:partitioning}
Below, we formally present the theoretical framework and assumptions 
that are common to all the three methods detailed in the next sections.
In this paper, we describe the many-body electron dynamics at the level of TDDFT~\cite{RG84, Fun12}.
In this context, the electronic density of a 
many-body system 
\begin{equation*}
  \rho(\vecr,t) = \sum\limits_{i=1}^{N}|\varphi_i(\vecr;t)|^2\:,
\end{equation*}
is obtained from an auxiliary one -- the Kohn-Sham (KS) system -- of non-interacting 
fermions which wavefunction is represented by a single Slater determinant $\Psi(\vecr;t)$ 
composed of $N$ orbitals $\varphi_i(\vecr;t)$. These orbitals satisfy the 
time-dependent KS equations
\begin{equation}
  \imath\partial_t\varphi_i(\vecr;t) 
  = \hat{H}_\mathrm{KS}\,\varphi_i(\vecr;t)\:,
\end{equation}
with the time-dependent KS~Hamiltonian
\begin{equation}\label{eq:HKS}
  \hat{H}_\mathrm{KS}[\rho](\vecr;t) = -\frac{\Delta}{2}+V_\mathrm{KS}[\rho](\vecr;t)\:,
\end{equation}
and the time-dependent KS~potential
\begin{equation}
  V_\mathrm{KS}[\rho](\vecr;t)
  =V_\mathrm{ext}(\vecr;t)+\int d^3r'\,\frac{\rho(\vecr';t)}{|\vecr-\vecr'|}
  + V_\mathrm{xc}[\rho](\vecr;t)\:,
\end{equation}
composed by the external field of the ions and the laser field, 
the classical Hartree and the exchange and correlation (xc) potential.
Once the time-dependent density is obtained by solving these equations, it is 
in principle possible to access any kind of observable provided it is expressed as a functional 
of $\rho(\vecr,t)$.

The momentum probability distribution of emitted electrons $P(\veck)$, i.e. 
the probability to measure an electron with momentum $\veck$ at a detector positioned far away from 
a target system, is the observable we aim to describe. 
In this work, we focus on the formulations that can be applied to real-space implementations 
and that require the knowledge of the wavefunction in a limited volume. 
All these approaches are resting on two principal assumptions.

\begin{figure}
  \centering
  \includegraphics[width=0.35\linewidth]{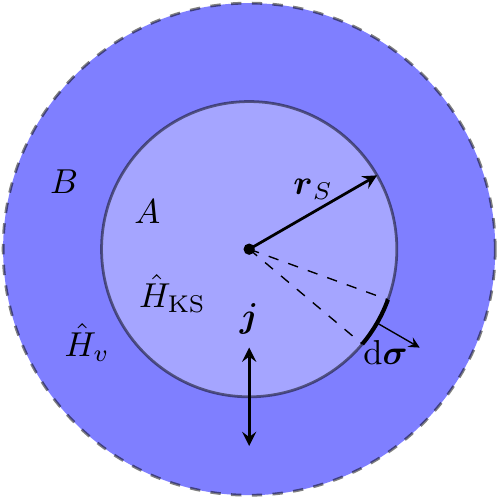}
  \caption{
  Schematic description of space partitioning in the surface flux, sampling point, 
  and mask methods. A surface $S$ (here chosen as spherical) separates the 
  space into region $A$ which is described by TDDFT and the Volkov propagation region $B$.}
  \label{fig:sketch-tsurff}
\end{figure}

The first assumption is that the dynamics of ionization can be accurately described by 
two different Hamiltonians localized in adjacent spatial regions $A$ and $B$  
separated by a surface $S$ as in Fig.~\ref{fig:sketch-tsurff} -- we here choose a spherical surface 
of radius $r_S$, but the shape can be general.
More specifically, we assume that in $A$ the electrons can be described with the KS Hamiltonian of 
Eq.~\ref{eq:HKS} while in $B$ they follow the exactly solvable Volkov Hamiltonian 
\begin{eqnarray}\label{eq:Hpartition}
  \hat{H}^B(t) = \hat{H}_v(t) = \frac{1}{2}\left[-\imath\nabla-\frac{\veca(t)}{c}\right]^2
  \label{eq:volkovhamiltonian}
\end{eqnarray}
governing the time evolution of free electrons in an external field $\veca(t)$. 
This means that the total time-dependent Hamiltonian $\hat{H}(\vecr;t)$ describing our system 
can be expressed as a combination of these two spatially localized Hamitonians,
\begin{equation}\label{eq:H_HAHB}
   \hat{H}(\vecr;t)=\left\{
    \begin{array}{l}
      \hat{H}_A(\vecr;t)  = \hat{H}_\mathrm{KS}(\vecr;t)  \quad \vecr\in A \\
      \hat{H}_B(\vecr ;t) = \hat{H}_v(\vecr;t)  \quad\,\,\, \vecr\in B
    \end{array} 
    \right. \,,
\end{equation}
and that electrons in $B$ can be safely described as free, independent particles.
The quality of such an assumption is limited by 
the error in the truncation of the tail of the Coulomb potential and ultimately 
depends on the position of $S$ relative to the target system. 
While this is exact in the proximity of the detector -- at an infinite distance -- and 
certainly bad in the vicinity of the system -- where electrons are strongly interacting with 
each other and with the external potential -- its 
range of applicability in many relevant situations is quite large. 
In practice, however, one has to converge the final results with respect to $S$.

Under the assumption that the space partitioning is a good approximation, we can 
represent the wavefunction in $B$ with a KS Slater determinant $\Psi^B(\vecr;t)$  
yet expand each orbital $\varphi_i^B(\vecr;t)$ on Volkov waves as follows
\begin{eqnarray}
  \varphi_i^B(\vecr;t) = \int d^3k\, b_{i}(\veck)\chi_\veck(\vecr;t)\:,
  \label{eq:expansion}
\end{eqnarray}
where
\begin{eqnarray}  \label{eq:volkovstate}
  \chi_\veck(\vecr;t) = (2\pi)^{-3/2}e^{-\imath\Phi(\veck;t)}e^{\imath\veck\vecr}\:,\quad {\rm with }\quad
  \Phi(\veck;t) = \frac{1}{2}\int\limits_0^t\,d\tau\left[\veck-\frac{\veca(\tau)}{c}\right]^2\:
\end{eqnarray}
are the exact solutions of the time-dependent Schr\"odinger equation for the
Volkov Hamiltonian in Eq.~\ref{eq:volkovhamiltonian}, $\Phi(\veck;t)$ is the Volkov phase 
and $b_i(\veck)$ is the spectral amplitude of the $i$th KS orbital.
Volkov waves are essentially plane waves with an additional time-dependent phase.

The second assumption is that in the long time limit, for $t\geq T$,
bound and ionized electronic wavefunctions are well localized in $A$ and $B$,
respectively.
In other words, we demand that the total density is given by $\rho(\vecr;T) = \rho^A + \rho^B$ with
\begin{eqnarray}
  \rho^A = \sum_{i=1}^{N} |\varphi_i^A(\vecr;T)|^2\:,\qquad
  \rho^B = \sum_{i=1}^{N} |\varphi_i^B(\vecr;T)|^2\:,
\end{eqnarray}
and with $\varphi_i^A(\vecr;T)$ ($\varphi_i^B(\vecr;T)$) being the $i$th KS~orbital localized in $A$ 
($B$) at time $T$.
This is exact for $T\rightarrow\infty$, while for finite values of $T$ one has to propagate for a time long 
enough such that slowest escaping electrons have reached region $B$.

Under the former assumptions the total number of electrons escaped from $A$ at time~$T$, 
which is precisely the integral of the total density in $B$,
can be expressed as
\begin{eqnarray}
  N_\mathrm{esc} = 
  \int\limits_B d^3r\,\rho(\vecr;T)=\int d^3r\,\rho^B(\vecr;T) = \sum_{i=1}^{N}\int d^3k\,|b_i(\veck)|^2
  \label{eq:nesc1}
\end{eqnarray}
where we used Eq.~\ref{eq:expansion} to obtain the rightmost expression.
Since the momentum distribution $P(\veck)$ is the differential probability  
associated with the total number of escaped electrons we can use the 
completeness of Volkov functions 
to invert Eq.~\ref{eq:expansion} and obtain that
\begin{eqnarray}\label{eq:pk_def}
  P(\veck) = \frac{d^3N_\mathrm{esc}}{d^3k}=\sum\limits_{i=1}^{N} |b_i(\veck)|^2 =\sum\limits_{i=1}^{N} 
   \left|\langle\chi_\veck(T)|\varphi_i^B(T)\rangle\right|^2\:.
\end{eqnarray}
From this equation it is apparent how $P(\veck)$ connects with the spectral amplitude 
$b_i(\veck)$ of the scattering orbitals expanded on Volkov waves. 
Less resolved quantities, like the energy-resolved photoelectron probability 
$P(\ekin =k^2/2)$,  can be obtained from $P(\veck)$ by direct integration.

The three methods which we describe in the next sections, essentially provide different 
approaches to obtain $b_i(\veck)$ from the knowledge of the KS orbitals in $A$ only.
Our aim is to have an expression for $b_i(\veck)$ which can be efficiently implemented in 
a real space three dimensional representation, that is accurate from linear to strong-field 
regimes and with an energy resolution comparable to experiment ($\Delta E \sim 0.05$ eV).
As we will show in the following, the SPM is straightforward to implement efficiently and 
provides a large flexibility in reciprocal space. However, it is unreliable 
in the mid- to strong-field regimes where it requires the use of unpractically 
large simulation boxes to properly converge under its working assumptions. 
MM, in contrast, yields reliable and precise results in all regimes, however, 
a computationally efficient implementation is more involved as Fourier transforms are needed. 
Moreover, when using this method the reciprocal space grid is constrained by the choice of the real-space 
one and spectra at low kinetic energies are difficult to retrieve.
\tsurff\ can be derived in a TDDFT formalism without any additional assumptions. 
The grid in momentum space can be chosen arbitrarily which allows the calculation of 
PES and PAD up to high precision and resolution.
Furthermore, by introducing an additional parameter (the angular momentum cut-off $L_\mathrm{max}$) 
the method can be efficiently implemented in three dimensions. 

\subsection{Sampling point method}\label{sec:spm}
The sampling point method (SPM) is the oldest and least accurate amongst the methods we discuss 
in this work~\cite{Poh00}. It was first presented without formal theoretical derivation and 
justified only on the basis of its results on practical calculations. We hereby present a 
derivation with particular 
attention to the conditions under which it is supposed to work reliably.

The SPM is based on the idea that $P(\veck)$ can be calculated by the simple knowledge of the 
time dependence of each orbital sampled only at a single point $\vecr_S$ of the surface $S$ in 
Fig.~\ref{fig:sketch-tsurff}. Besides the assumptions discussed in the previous section, it rests 
on the additional conditions that \textit{(i) }$\vecr_S$ is positioned at a sufficiently large 
distance such that the ionized wavepackets arrive when the laser pulse has been switched off 
(for $t\geq T_\mathrm{pulse}$) and \textit{(ii)} with a final momentum directed along 
$\vecr_S$, i.e., $\veck=k\vece_{\vecr_S}$.

Under these conditions we can drop the field $\mathbf{A}(t)$ in the Volkov phase which then 
describes free particles, and explicitly write the expansion of Eq.~\ref{eq:expansion} as
\begin{eqnarray}
  \varphi_i^{B}(\vecr_S;t ) 
  = (2\pi)^{-3/2}\int d\veck\, b_i(k,\Omega_{\vecr_S})e^{\imath kr_S-\imath k^2t/2}\:,
  \label{eq:sampling}
\end{eqnarray}
where we express the momentum vector in spherical coordinates $\veck=(k,\Omega_{\vecr_S})$ to 
stress the form enforced to the final momentum by condition (ii). 
A Fourier transform in the time domain is then sufficient to impose the free particle dispersion 
relation, $E=k^2/2$, and extract the Volkov amplitudes with the following result
\begin{equation}\label{eq:teFT}
  \tilde{\varphi}_i^B(\vecr_S;E) = \frac{1}{\sqrt{2\pi}}\int dt\,e^{i E t}\varphi_i^B(\vecr_S;t ) =
  \frac{e^{\imath \sqrt{2E}r_S}}{\sqrt{2E(2\pi)^{3}}}\,b_i(\sqrt{2E},\Omega_{\vecr_S})\:.
\end{equation}
To obtain the above relation we used the time condition (i) and the 
Dirac delta $\delta(E-k^2/2)$ resulting from the time integral to simplify Eq.~\ref{eq:sampling}.
At this point the momentum distribution probability can be straightforwardly obtained from 
Eq.~\ref{eq:pk_def} as
\begin{equation}
  P\left(\veck =(\sqrt{2E}, \Omega_{\vecr_S})\right) = 
  2 E(2\pi)^3  \sum_{i=1}^{N}\left|\tilde{\varphi}_i^B(\vecr_S;E)\right|^2\:,
  \label{eq:raw}
\end{equation}
where we explicitly inverted the dispersion relation to obtain the 
momentum magnitude $k=\sqrt{2E}$. This implies that $k$ is always positive and therefore we must 
further
impose that \textit{(iii)} at the sampling point, the electrons are strictly outgoing.

The SPM working conditions are asymptotically valid for $\vecr_S$ positioned at an infinitely large distance 
from the system, but quickly degrade as we move closer. 
The most stringent condition is the time constraint (i) since it directly forces $\vecr_S$ to be 
positioned at a distance that proportionally grows with the laser switch-off time $T_{\rm pulse}$. 

A simple way to overcome this limitation was proposed in Ref.~\cite{Di12b}. 
It substantially reduces to keeping the full Volkov phase, including the field, in the expansion 
of Eq.~\ref{eq:expansion} and to compensate it in the Fourier time integral.
The Fourier exponent $i Et$ in Eq.~\ref{eq:teFT} is thus substituted with the Volkov phase 
$i \Phi(\veck;t)$ evaluated at $\veck=\sqrt{2E}\vece_{\vecr_S}$, 
\begin{equation}
  \tilde{\xi}_i^B(\vecr_S;E) = \frac{1}{\sqrt{2\pi}}\int dt\,
 e^{\imath\Phi(\sqrt{2E}\vece_{\vecr_S};t)}\,\varphi_i^B(\vecr_S;t)\:.
\end{equation}
The photoelectron momentum distribution is then obtained by simply replacing 
$\tilde{\varphi}_i^B(\vecr_S;E)$ with $\tilde{\xi}_i^B(\vecr_S;E)$ in Eq.~\ref{eq:raw}.
Owing to the presence of the Volkov phase in the time integral this variant
goes under the name of phase-augmented sampling point method (PA-SPM).

Even though this approach is superior to the simple SPM, it is still limited by 
conditions (ii) and (iii). The validity of these conditions is difficult to assess in practical 
calculations since it strongly depends on the electron dynamics induced by the external field and 
can only be
taken under control by converging the final results with respect to the position of $\vecr_S$.

\subsection{Surface flux method}\label{sec:sfm}
In contrast to the SPM, the time-dependent surface flux method~(\tsurff)~\cite{Tao12,Scr12,Cai05}
makes no further assumption besides the ones discussed in Sec.~\ref{sec:partitioning}.
Thus, for instance, it can handle situations where electrons are driven 
by the laser field back towards the emission site like in the backscattering regime.
We here describe a derivation alternative to the one present in the literature.
Our derivation is 
based on the flux of the current-density operator through $S$~(see the scheme of 
Fig.~\ref{fig:sketch-tsurff}) that is suitable for TDDFT.

Owing to the space and Hamiltonian partitioning explained in Sec.~\ref{sec:partitioning}, 
we can describe the electronic wavefunctions with both $\Psi_A(\vecr;t)$ and $\Psi_B(\vecr;t)$
on the surface $S$ that separates region $A$ and $B$. 
Using the continuity equation we thus express the total number of escaped electron $N_{\rm esc}$ at time $T$ 
in terms of the flux integral 
\begin{eqnarray}
  N_\mathrm{esc} &=  \int\limits_0^T dt\int\limits_B d^3r\,\frac{d\rho(\vecr;t)}{dt}
  = -\int\limits_0^T dt\int\limits_{S}d\mbox{\boldmath$\sigma$} \cdot\mbraket{\Psi^B}{\hat{\mathbf{j}}}{\Psi^A}\nonumber\\
  &=-\sum_{i=1}^{N}\int\limits_0^T dt\int\limits_{S}
   d\mbox{\boldmath$\sigma$}\cdot\mbraket{\varphi_i^B}{\hat{\mathbf{j}}_i}{\varphi_i^A}
  \label{eq:nesc3}
\end{eqnarray}
of the single-particle, gauge-invariant, current-density operator
\begin{equation}
  \hat{\mathbf{j}}_i(\vecr) = 
  \frac{1}{2\imath}\left[
    \left(\nabla_i-\imath\frac{\veca}{c}\right)\delta(\vecr-\vecr_i)+
    \delta(\vecr-\vecr_i)\left(\nabla_i
  -\imath\frac{\veca}{c}\right)\right]\:,
\end{equation}
evaluated over $\Psi_A(\vecr;t)$ and $\Psi_B(\vecr;t)$ or the orbitals which they are composed of. 
We then replace the bra in Eq.~\ref{eq:nesc3} and insert the expansion Eq.~\ref{eq:expansion} in Volkov states
\begin{eqnarray}
  N_\mathrm{esc}
  = -\sum_{i=1}^{N}\int\limits_0^T dt\int\limits_{S}d\mbox{\boldmath$\sigma$}\cdot
  \int d^3k\left(b^*_i(\veck)\mathbf{J}^{(i)}_\veck\right)
  \label{eq:nesc2}
\end{eqnarray}
with $\mathbf{J}_\veck^{(i)}\equiv\mbraket{\chi_\veck}{\hat{\mathbf{j}}_i}{\varphi_i^A}$.
Since the choice of the subscript $A$ and $B$ in the brakets of Eq.~\ref{eq:nesc3} is arbitrary, 
we can equivalently choose the opposite order and obtain that $N_\mathrm{esc}$ is also equal to 
Eq.~\ref{eq:nesc2} complex conjugated.
Comparing Eqs.~\ref{eq:nesc1} with \ref{eq:nesc2} and its complex conjugated for each 
single orbital yields the final expression for the spectral amplitude
\begin{eqnarray}
  b_i(\veck) = -\int\limits_0^T dt\int\limits_{S}d\mbox{\boldmath$\sigma$}
                \cdot\mathbf{J}_\veck^{(i)}\:,
  \label{eq:flux}
\end{eqnarray}
in terms of the Volkov projected single-particle current density $\mathbf{J}_\veck^{(i)}$.
The momentum probability distribution $P(\veck)$ can then be obtained from 
Eq.~\ref{eq:pk_def} by summing up $|b_i(\veck)|^2$ over the orbital index $i$.

From Eq.~\ref{eq:flux} it is apparent that the extension to TDDFT is straightforward.
In practical implementations one needs to calculate $\mathbf{J}^{(i)}_\veck$ for a given set 
of $\veck$ and accumulate its flux integral over time. 
To this end, one needs only to keep track of the KS orbitals $\varphi_i^A(\vecr;t)$ and 
their gradients over $S$ while the Volkov waves (and their gradients) are analytical. 
In principle, provided $S$ is positioned far enough from the system, there is no restriction 
to the choice of its shape. 
However, we found that a spherical surface is advantageous from the numerical standpoint as 
it allows to expand the Volkov waves in spherical harmonics to decouple $\veck$ and $\vecr$.  
This in turn, requires to truncate the integrals over the sphere
up to a given maximum angular momentum $L_\mathrm{max}$ and thus introduces  
an additional parameter to converge (see Appendix~A).
In practice, we observed that $L_\mathrm{max}\approx 100$ is enough for a large class of problems
involving moderately strong fields ($I\lesssim 10^{14}\,\wcmq$).

Finally, we mention that a variant of the sampling point method can be derived by truncating 
the surface integral in Eq.~\ref{eq:flux} to a single point (see Appendix~B). 
This leads to an alternative expression similar to the one for PA-SPM.
In our tests, however, we found that this variant did not present any significant improvement 
over PA-SPM and therefore we did not develop it further.

\subsection{Mask method}\label{sec:mm}
Similar in philosophy to \tsurff, the mask method~(MM) is derived under the same 
assumption on the ionization process. We here recall the salient traits and remind the 
reader to Ref.~\cite{deGio12,DeGio13,CrU14b} for further details.  

As discussed in Sect.~\ref{sec:partitioning}, in the long-time limit of an ionization 
process, we can assume that the electronic density and hence the wavefunction 
splits into two spatially separated parts. 
A practical way to implement this splitting for a generic time $t$ is to use a mask 
function~$M(\mathbf{r})$ on each KS orbital as follows
\begin{equation}\label{eq:mask_partition}
  \varphi_i(\mathbf{r};t) 
  = M(\mathbf{r})\varphi_i(\mathbf{r};t) +[1-M(\mathbf{r})] \varphi_i(\mathbf{r};t)
  = \varphi_i^A(\mathbf{r};t)+\varphi_i^B(\mathbf{r};t)\:,
\end{equation}
where $M(\mathbf{r})$ is a continuous function equal to 1 in the inner part of $A$ 
and that smoothly decays to 0 in $B$.

Using the the mask we can formally write the solution of the TDKS equations in the whole space 
as a set of coupled equations,
\begin{eqnarray}\label{eq:MM_split_1}
   \left\{
    \begin{array}{l}
      |\Psi^A(t^\prime)\rangle = \hat{M}\hat{U}(t^\prime,t)\left[ |\Psi^A(t)\rangle + |\Psi^B(t)\rangle\right]  \\
      |\Psi^B(t^\prime)\rangle = [1-\hat{M}]\hat{U}(t^\prime,t)\left[ |\Psi^A(t)\rangle + |\Psi^B(t)\rangle\right] 
    \end{array}
   \right.\,,
\end{eqnarray}
using the time evolution operator
\begin{equation}\label{eq:MM_UT}
  \hat{U}(t^\prime,t)= \exp\left\{ -i \int_{t}^{t^\prime} \hat{H}(\tau) d \tau \right\}\,,
\end{equation}
with the time-boundary condition $|\Psi^B(t=0)\rangle=0$ and with a mask operator 
defined as $\hat{M} = \sum_{i=1}M(\mathbf{r}_i)\delta(\mathbf{r}_i-\mathbf{r}'_i)$.

Owing to the asymptotic condition Eq.~\ref{eq:volkovhamiltonian} on the Hamiltonian, 
$|\Psi^B(t)\rangle$ evolves under the action of  $\hat{H}_v$, 
and we indicate with $U_v(t^\prime,t)$ the associated evolution operator.
Since $\hat{H}_v$ is diagonal in momentum and $\hat{H}_{\rm KS}$ 
is almost local in real space, we can write the equation of motion in a mixed real and 
momentum space representation.
In this representation we can integrate Eq.~\ref{eq:MM_split_1} by 
recursively applying the discrete time evolution operator 
$\hat{U}(\Delta t)\equiv\hat{U}(t+\Delta t,t)$ as follows
\begin{eqnarray}\label{eq:FMM_prop}\fl\quad
   \left\{
    \begin{array}{l}
      \langle \mathbf{r}|\Psi^A(t+\Delta t)\rangle = 
         \langle \mathbf{r}|\hat{M}\hat{U}(\Delta t)|\Psi^A(t)\rangle + \langle 
      \mathbf{r}|\hat{M}\hat{U}_v(\Delta t)|\Psi^B(t)\rangle  \\
      \langle \chi_\veck|\Psi^B(t+\Delta t)\rangle = 
        \langle \chi_\veck|[1-\hat{M}]\hat{U}(\Delta t)|\Psi^A(t)\rangle +   
      \langle \chi_\veck|[1-\hat{M}]\hat{U}_v(\Delta t)|\Psi^B(t)\rangle
    \end{array}
   \right.\,,
\end{eqnarray}
with the initial condition $\langle \chi_\veck|\Psi^B(t=0)\rangle=0$.
These equations can be written in a closed form for $\langle \mathbf{r}|\Psi^A(t)\rangle$ 
and $\langle \chi_\veck|\Psi^B(t)\rangle$, 
by including the following set of equations, here explicitly expanded for 
each KS orbital
\begin{eqnarray}\label{eq:FMM_prop_aux}\fl\quad
  \left\{
  \begin{array}{l}
  \langle \mathbf{r}|\hat{M}\hat{U}(\Delta t)|\varphi_i^A(t)\rangle =  M(\mathbf{r})
  \langle \mathbf{r}|\hat{U}(\Delta t)|\varphi_i^A(t)\rangle\\
  \langle \mathbf{r}|\hat{M}\hat{U}_v(\Delta t)|\varphi_i^B(t)\rangle =
  M(\mathbf{r}) \int  \langle \mathbf{r}|\chi_\veck\rangle 
  \langle \chi_\veck|\varphi_i^B(t)\rangle d^3 k\\
  \langle \chi_\veck|[1-\hat{M}]\hat{U}(\Delta t)|\varphi_i^A(t)\rangle =
   \int  \langle \chi_\veck|\mathbf{r}\rangle [1-M(\mathbf{r})] 
  \langle \mathbf{r}|\hat{U}(\Delta t)|\varphi_i^A(t)\rangle d^3 r\\
  \langle \chi_\veck|[1-\hat{M}]\hat{U}_v(\Delta t)|\varphi_i^B(t)\rangle =
  \langle \chi_\veck|\varphi_i^B(t)\rangle -
  \int \langle \chi_\veck|\mathbf{r}\rangle
  \langle \mathbf{r}|\hat{M}\hat{U}_v(\Delta t)|\varphi_i^B(t)\rangle d^3 r
  \end{array}
  \right. .
\end{eqnarray}

Once Eqs.~\ref{eq:FMM_prop} and \ref{eq:FMM_prop_aux} are propagated up to 
time $T$, 
the momentum distribution is straightforwardly obtained by summing up the square modulus of the 
KS orbitals as in Eq.~\ref{eq:pk_def}, namely: $ P(\veck)=\sum_{i=1}^{{N}}|\langle \chi_\veck|\varphi^B_{i}(T)\rangle|^2$.
Unlike the approaches described in the previous sections, since Eqs.~\ref{eq:FMM_prop} and 
\ref{eq:FMM_prop_aux} include the boundary conditions for the wavefunctions in $A$ and $B$, there 
is no need for additional absorbing boundaries.

In a numerical implementation the evaluations of the integrals in Eq.~\ref{eq:FMM_prop_aux}
must undergo some level of discretization.
In particular, substituting Fourier integrals with Fourier series introduces unwanted periodic 
boundaries conditions that reintroduce ionized wavepackets into the simulation box and eventually 
lead to instability (for details see the appendix of Ref.~\cite{deGio12}).

A stabler scheme can be obtained by simplifying Eq.~\ref{eq:FMM_prop_aux} under the 
assumption that the electron flow 
is only outward from $A$.  
In this case we can omit the term responsible for the introduction of 
charge from $B$, and 
obtain the modified set of equations
\begin{eqnarray}\label{eq:MM_prop_aux}\fl\quad
  \left\{
  \begin{array}{l}
  \langle \mathbf{r}|\hat{M}\hat{U}(\Delta t)|\varphi_i^A(t)\rangle =  M(\mathbf{r})
  \langle \mathbf{r}|\hat{U}(\Delta t)|\varphi_i^A(t)\rangle\\
  \langle \mathbf{r}|\hat{M}\hat{U}_v(\Delta t)|\varphi_i^B(t)\rangle = 0 \\
  \langle \chi_\veck|[1-\hat{M}]\hat{U}(\Delta t)|\varphi_i^A(t)\rangle =
   \int \langle \chi_\veck|\mathbf{r}\rangle [1-M(\mathbf{r})] 
  \langle \mathbf{r}|\hat{U}(\Delta t)|\varphi_i^A(t)\rangle d^3 r\\
  \langle \chi_\veck|[1-\hat{M}]\hat{U}_v(\Delta t)|\varphi_i^B(t)\rangle =
  \langle \chi_\veck|\varphi_i^B(t)\rangle   \end{array}
  \right. .
\end{eqnarray}
Together with Eq.~\ref{eq:FMM_prop} it defines a modified scheme completely 
equivalent to the previous one in the limit 
where $r_S$ is big enough to justify the outgoing flow condition.
We note that, compared to Eq.~\ref{eq:FMM_prop_aux}, the first two equations in 
Eq.~\ref{eq:MM_prop_aux} governing the evolution of the real-space components of the 
wavefunction in $A$, are no longer connected with the momentum-space ones. 
For this reason the propagation is thus equivalent to a time propagation with a mask function absorber that can introduce spurious reflections at the boundaries. 
Such reflections can, in principle, be reduced by using the most appropriate 
mask function absorber or a complex absorbing potential casted in the form of a mask function~\cite{deGio14}.
In the energy range where the mask function absorbs well, it is possible to 
carry out stable simulations for long times.

\section{Examples}\label{sec:examples}
In the following, we illustrate the above mentioned approaches
with a few examples. 
Unless otherwise specified, we use TDDFT at the level of the 
 time-dependent (adiabatic) local-density approximation 
(ALDA)~\cite{PW92}, 
augmented by an average-density self-interaction correction~(SIC)~\cite{Leg02}
which corrects the tail of the Coulomb potential and yields an accurate ionization potential.
Furthermore, in order to prevent artificial reflections at the borders of the simulation 
box we employ absorbing boundary conditions.
In all the simulations the ions were clamped to their equilibrium positions.

All numerical calculations were performed with the
real-time, real-space TDDFT code \textsc{Octopus}
freely available under the GNU public license~\cite{And15, Cas06}.

\subsection{Hydrogen atom}
We here present a comparison of all the methods discussed in this paper. 
To this end, we choose as a benchmark test the case of above-threshold ionization~(ATI) in an 
hydrogen atom. 
Clearly, there is no need to use TDDFT for a one-electron system, and our interest here is 
focused to assess the numerical performance and the accuracy of the different methods. 
For this reason the simulations were carried out at the level of single-particle TDSE.

We choose a Cartesian grid of spherical shape with radius 
$\rbox =90\,$a.u.~including an outer 
shell of width 40\,a.u.\ with a complex absorbing potential of height 
$\eta = -0.2$~\cite{deGio14}.
We employ a pulse of $N_c=20$ cycles, linearly polarized along the $z$-axis with 
wavelength $\lambda = 800\,$nm~($\omlas = 1.55$\,eV), and intensity $I=5\times 10^{13}\,\wcmq$.
Photoelectrons are collected until shortly after the pulse, where the total ionization amounts to 
$\nesc \sim 10^{-3}$.
The surface points for the \tsurff\ method and the sampling points for the 
SPM are both located on a 
sphere of radius $r_S=50\,$a.u.\ directly in front of the absorbing zone.
The flux is evaluated with the expansion in spherical harmonics, Eq.~\ref{eq:sphflux}, 
up to a maximum angular momentum~$L_\mathrm{max}=20$. The data 
of the MM is extracted from Ref.~\cite{deGio12} where a box of radius $\rbox=60\,$a.u.\ 
and a mask absorber of width 10\,a.u.~was used.

\begin{figure}
  \centering
  \includegraphics{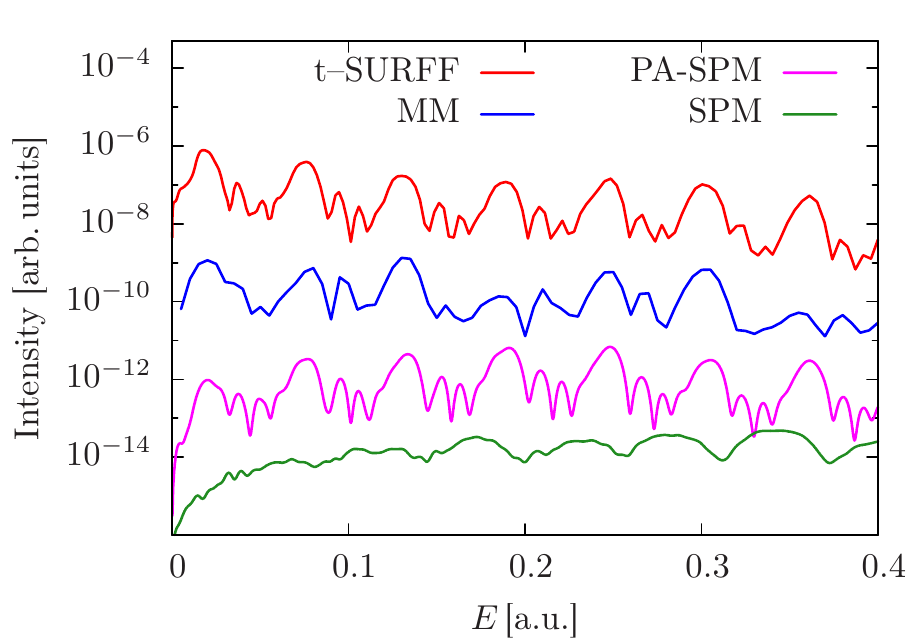}
  \caption{
    Calculated photoelectron spectra of hydrogen exposed to a 20-optical-cycle laser pulse
  with a peak intensity of $I = 5\times 10^{13}\,\wcmq$ and a frequency of 
  $\omlas  = 0.057$\,a.u., obtained with the \tsurff, 
  mask (MM), and sampling 
  point methods~(SPM and PA-SPM).
  }
  \label{fig:hydrogen-pes}
\end{figure}
Figure~\ref{fig:hydrogen-pes} shows the total spectra obtained with the three different 
methods. Apart from the SPM, all methods compare quite well and display a series of 
ATI peaks separated by the photon energy $\omlas = 0.057\,$a.u.
In contrast, the SPM presents a featureless background for low-energies which, as the energy
increases, transforms into a series of peaks roughly spaced by $\omlas$, but with the wrong onset.
For this reason, we conclude that the 
SPM is not suitable for laser excitations in this regime ($\gamma = 1.5$). 

\begin{figure}
  \centering
  \includegraphics[width=1.0\textwidth]{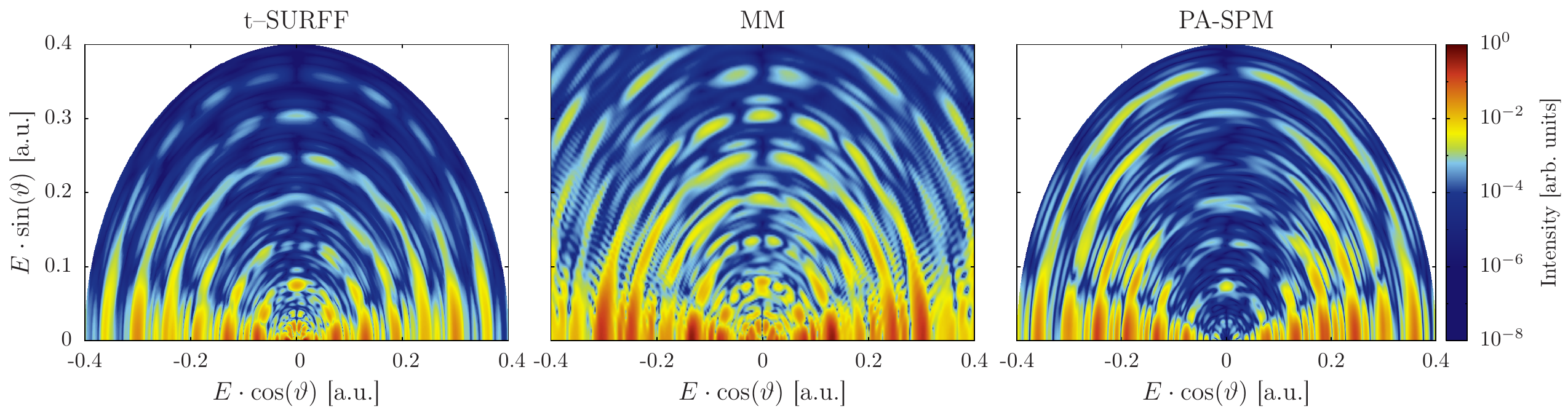}
  \caption{
    Calculated photoelectron energy-angular distributions of hydrogen exposed to 
    a 20-optical-cycle laser pulse with 
    a peak intensity of $I=5\times 10^{13}\,\wcmq$ and a frequency of $\omlas = 0.057\,$a.u.,
    obtained with the surface flux (left panel), mask (middle panel), 
    and (phase-augmented) sampling point methods (right panel).}
  \label{fig:hydrogen-pad}
\end{figure}

The different quality of the results can be better assessed from the 
angle-resolved spectra. Figure~\ref{fig:hydrogen-pad} displays high-resolution density plots of the 
spectra as a function of the kinetic energy $\ekin$ and the angle $\vartheta$ measured 
with respect to the laser polarization axis, obtained with \tsurff~(left), 
MM (middle), and PA-SPM (right). 

The ATI peaks unfold into rings with a number of stripes 
equal to the angular momentum quantum number of the dominant partial 
wave in the final state plus one~\cite{Arb06}. The low-energy region shows a peculiar nodal 
pattern which is induced by the long-range Coulomb potential.
The pattern for the \tsurff~method compares very well with the MM
and with similar calculations in the literature~\cite{Zhou11}. 
This demonstrates that \tsurff\ is 
indeed a reliable tool to calculate photoelectron energy-angular distributions in this regime.

Comparing \tsurff~method and 
MM to PA-SPM, we observe significant differences. 
While the emission is preferentially along the laser polarization in all cases, PA-SPM
underestimates the emission in other directions. In particular, it hardly reproduces the 
stripes perpendicular to the laser polarization and the low-energy region which is 
sensitive to the tail of the Coulomb potential.
This failure suggests that the contribution of electrons with momentum not parallel to $\vecr_{S}$
neglected by PA-SPM is crucial to form these interference patterns. 
Therefore, we can conclude that only \tsurff\ and MM can be recommended 
for mid- to strong-field regimes and that all variants of SPM should be avoided.
Finally, we mention that the single-point approximation of the \tsurff~method~(see Appendix~B) 
yields results (not shown) which are in line with those obtained with PA-SPM.

\subsection{C$_{60}$ fullerene}
Compared to the previous section, we here tackle the more challenging problem in the 
rescattering 
regime ($\gamma < 1$) where the use of TDDFT is mandatory.
Following~Ref.~\cite{Gao16}, we consider the fullerene 
C$_{60}$ exposed to a strong laser pulse, linearly polarized along the $z$-axis with 
frequency $\omlas=1.36\,$eV, intensity 
$I=10^{14}\,\wcmq$, and pulse length $\tpulse = 24\,$fs.
In Ref.~\cite{Gao16} the positively charged ionic background of the molecule was 
approximated by a jellium 
shell and calculations where performed in cylindrical coordinates. 
As already mentioned in the reference, this
model suffers from the fact that returning electrons collide with a jellium well instead of a 
carbon ion
which eventually leads to an underestimation of high-energy electrons.
This is a limitation since the angular pattern of rescattered electrons is actually influenced  
by the interatomic distances as it results from the interference of waves scattering  
from different ionic centers~\cite{Lin10, Xu10}. This is crucial, for instance, for laser-induced 
electron diffraction~\cite{Mec08, Bla12}. Therefore, we choose a three dimensional description which  
includes the ionic background by modeling each atom with a pseudo-potential~\cite{TM91}. 
In what follows we consider only a single orientation of the molecule relative
to the laser polarization.

\begin{figure}
  \centering
  \includegraphics[width=0.5\linewidth]{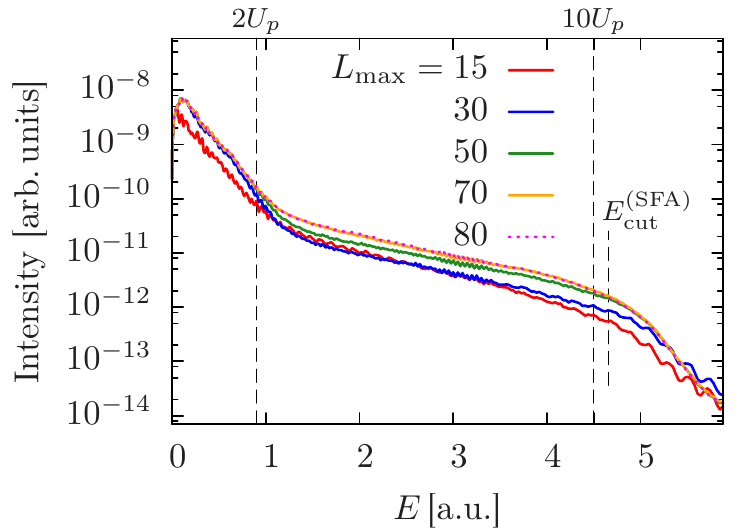}
  \caption{Calculated photoelectron spectra of C$_{60}$ exposed to a fs pulse with a peak 
  intensity of 
  $I=1.6\times 10^{14}\,\wcmq$ and a frequency of $\omlas = 0.05\,$a.u., obtained for different 
  cut-offs $L_\mathrm{max}$
  in the expansion in spherical harmonics Eq.~\ref{eq:sphflux}.}
  \label{fig:c60-converge}
\end{figure}

\tsurff\ allows for computational boxes of the order of the free electron
quiver amplitude which for our laser is 
$x_p=27\,$a.u. Here we used $\rbox=70\,$a.u.~with the surface 
being located at $r_S=50\,$a.u.~in front of a complex absorbing potential 
of width 20\,a.u. and height $\eta = -1.0$.
A crucial parameter of the \tsurff\ implementation is the angular 
momentum cut-off $L_\mathrm{max}$. 
Figure~\ref{fig:c60-converge} shows the obtained spectra for different $L_\mathrm{max}$.
From the figure it is clear that 
convergence is obtained first for the direct electrons ($E<2U_p$) at $L_\mathrm{max}\sim 30$, 
and only later, for much higher $L_\mathrm{max}\sim 80$, in the plateau region. 
Nevertheless, all spectra show a large plateau in the range of 
$\ekin=30\-- 125\,$eV up to the cut-off located at around 
$E_\mathrm{cut}^{\mathrm{(SFA)}}=10.007U_p+0.538\,E_\mathrm{IP}$. 

\begin{figure}
  \centering
  \includegraphics[width=0.8\linewidth]{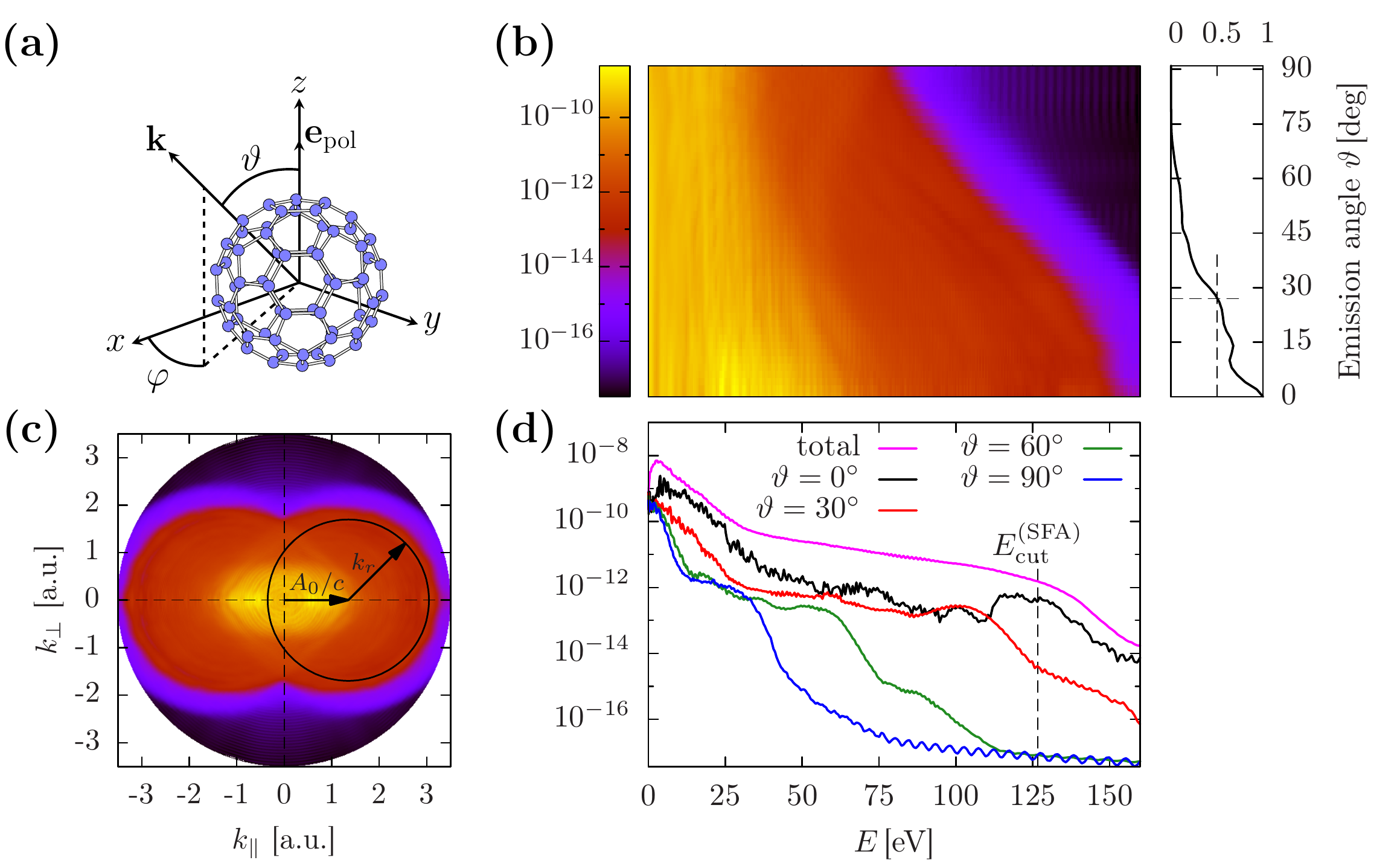}
  \caption{\textbf{(a)}~Geometry of C$_{60}$ fullerene and angles $(\vartheta,\varphi)$ 
  of the outgoing electrons measured with respect to the laser polarization axis 
  (along the $z$-axis). 
  \textbf{(b)}~Color map of calculated angle-resolved photoelectron spectrum 
  of C$_{60}$ as a function of 
  the kinetic energy $\ekin$ and emission angle $\vartheta$, in logarithmic scale. 
  The full three-dimensional momentum distribution was
  averaged over the angle $\varphi$. 
  The right part of (b) shows the PAD obtained from the integration of the angular-resolved PES over 
  the high-energy range of 50--160\,eV, normalized to 1 at $0^\circ$.
  The laser pulse parameters are: $\omlas=1.36\,$eV, $\tpulse=24\,$fs, $I=1.6\times 10^{14}\wcmq$.
  \textbf{(c)}~Similar to (b), but as a function of the momenta $k_\parallel$ and 
  $k_\perp$ parallel and perpendicular to the laser polarization, respectively.
  \textbf{(d)} Photoelectron spectra for different emission angles $\vartheta$ and 
  total PES (in logarithmic scale).}
  \label{fig:c60-arpes}
\end{figure}

Angular-resolved quantities are displayed in Fig.~\ref{fig:c60-arpes} ($L_\mathrm{max}=80$).
The angles $\vartheta$ and $\varphi$ are measured with respect to the laser polarization axis, 
see Fig.~\ref{fig:c60-arpes}~(a). The left part of Fig.~\ref{fig:c60-arpes}~(b) shows the full 
angular-resolved PES averaged over the angle $\varphi$, while in the right part we find 
the PAD obtained from an integration of the angular-resolved PES over the 
high-energy range of 50--160\,eV.
As one can see, in this energy range, almost all photoelectrons are emitted with 
an angle $\vartheta\leq 45^\circ$, i.e., in a cone in forward-backward direction. This in contrast to the 
jellium model, where the integrated PAD is strongly peaked 
around $\vartheta = 0^\circ$, we get a larger portion of electrons scattered sidewards. 
This confirms that the PAD of rescattered electrons is highly sensible to the ionic 
structure of the target. 

Figure~\ref{fig:c60-arpes}~(c)
displays the momentum distribution $P(\veck)$  
as a function of the momenta $k_\parallel$ and $k_\perp$ parallel and 
perpendicular 
to the laser polarization axis, respectively. The angular-resolved PES is now decomposed by 
two circles of 
maximum radii $k_r \approx 1.26A_0$ 
which are shifted by $\pm A_0/c$ with respect to 
the origin. 
This specific shape is the result of the rescattering process and characteristic
of the LIED regime. 
LIED features can be interpreted with the semianalytical models provided by the 
quantitative rescattering theory. 
According to the quantitative rescattering theory~\cite{Lin10, Xu10, Che09}, 
photoelectrons are released by tunnel ionization with 
an initial velocity of near zero. They then quiver in the laser field before returning back 
towards the target ion with incident momentum $\veck_0$ where they scatter elastically 
in all directions with scattered momentum $\veck_r$ ($|\veck_r|=|\veck_0|$).
The maximum kinetic energy that quiver electrons can gain in the laser field 
corresponds to $k_r^2/2 = 3.2\,U_p = 3.2/4\cdot A_0^2$, and 
since the elastic collision occurs in the laser field, photoelectrons 
gain an additional momentum $\veca(t_r)/c$ 
from the field at the recollision time $t_r$. The calculated PAD in Fig.~\ref{fig:c60-arpes}~(c) 
fits well with this model. 
The elastic scattering occurs in all directions. Thus, 
for a realistic simulation of strong-field ionization it 
is necessary to use an atomistic model which appropriately describes the rescattering process.

Figure~\ref{fig:c60-arpes}~(d) finally shows photoelectron spectra for different emission 
angles $\vartheta$. Electrons with highest kinetic energies are emitted exclusively
along the laser polarization axis, as can be seen at the line for $\vartheta = 0^\circ$. 
The cut-off is most clearly seen in the total spectrum and fits well with the theoretical 
prediction $E_\mathrm{cut}^\mathrm{(SFA)}$. Spectra of electrons that are emitted sidewards, also 
exhibit cut-offs, but with values smaller than $E_\mathrm{cut}^\mathrm{(SFA)}$.

\subsection{Organic molecules}
\label{sec:orbital_reconstruction}

In this section, we move from the strong-field regime to the linear one where the laser intensity is 
weak, still the photon fluence is large enough to justify the use of a classical description 
for the electromagnetic field. 

In this regime, electrons need to absorb only one photon to ionize. 
Much like in the photoelectric effect, electrons ejected in this regime carry 
information about the
energy level of their parent system encoded in the kinetic energy spectrum.
The spectrum is composed of a series of peaks positioned at kinetic energies 
$\ekin_i=\omega-E_{\rm IP}^{(i)}$ where 
$E_{\rm IP}^{(i)}$ is the ionization potential of the $i$th state of the system. 
This fact can be easily derived from time-dependent perturbation theory using Fermi's golden rule which,
apart from an inessential scaling factor, reads 
\begin{equation}\label{eq:fermi_gold}
  P(\mathbf{k})\propto \sum_i\left|\langle\Psi_f(\mathbf{k}) |\mathbf{A}\cdot \hat{\mathbf{p}} |\Psi_i\rangle \right|^2
  \delta(E_f-E_i-\omega)\,.
\end{equation}
This equation describes the probability to excite an electron from an initial state 
$|\Psi_i\rangle$ to $|\Psi_f\rangle$ separated by $E_{\rm IP}^{(i)}=E_f-E_i$ using an 
external field coupled with the dipole matrix element $\mathbf{A}\cdot\hat{\mathbf{p}}$
with $\hat{\mathbf{p}}$ being the momentum operator.

Using Eq.~\ref{eq:fermi_gold} as a starting point in Ref.~\cite{Puschnig:2009ho} it was first 
shown that 
photoelectrons carry also information about the orbitals from which they originate.
This information is encoded in the PAD, and can be isolated making the assumption that the 
final state is a plane wave $|\Psi_f(\mathbf{k})\rangle\approx|\mathbf{k}\rangle$, and that the 
initial state can be decomposed into separated orbitals $|\varphi_i\rangle$. 
Under these assumptions, and restricting to energies $E_i$ infinitesimally close to
$|\varphi_i\rangle$, Eq.~\ref{eq:fermi_gold} becomes 
\begin{equation}\label{eq:pk}
  P(\mathbf{k}) \propto 
    \left|\langle \mathbf{k} |\mathbf{A}\cdot \hat{\mathbf{p}} |\varphi_i\rangle \right|^2
  = |\mathbf{A}\cdot \mathbf{k}|^2 |\tilde{\varphi}_i(\mathbf{k})|^2 \,,
\end{equation}
where $\mathbf{k}$ is constrained to a spherical energy shell 
$E_i=\mathbf{k}^2/2=\omega-E_{\rm IP}^{(i)}$ and where
with $|\tilde{\varphi}_i(\mathbf{k})|^2$ we indicate the Fourier transform of the orbital.
Thus, apart from a purely geometrical factor $|\mathbf{A}\cdot \mathbf{k}|^2$, the angular 
distribution of photoelectrons turns out to be proportional to the Fourier transform of the parent 
orbital. 
The geometrical factor can be eliminated by summing up the PADs obtained with two perpendicular 
polarizations, for instance along $x$ and $y$,
\begin{equation}\label{eq:puk}\fl\quad
  P_U(\mathbf{k})=P_x(\mathbf{k})+P_y(\mathbf{k})\propto
  (|Ak_x|^2+ |Ak_y|^2)|\tilde{\varphi}_i(\mathbf{k})|^2=2 E_i^2A^2|\tilde{\varphi}_i(\mathbf{k})|^2\, ,
\end{equation}
and we obtain a direct connection between the Fourier transform of orbitals and photoelectron data.  
Combining incoherently PADs obtained with perpendicular polarizations is equivalent to the use of a 
single unpolarized pulse. For this reason we named $P_U(\mathbf{k})$ the result of the previous 
equation.

\begin{figure}
  \includegraphics[width=1.0\linewidth]{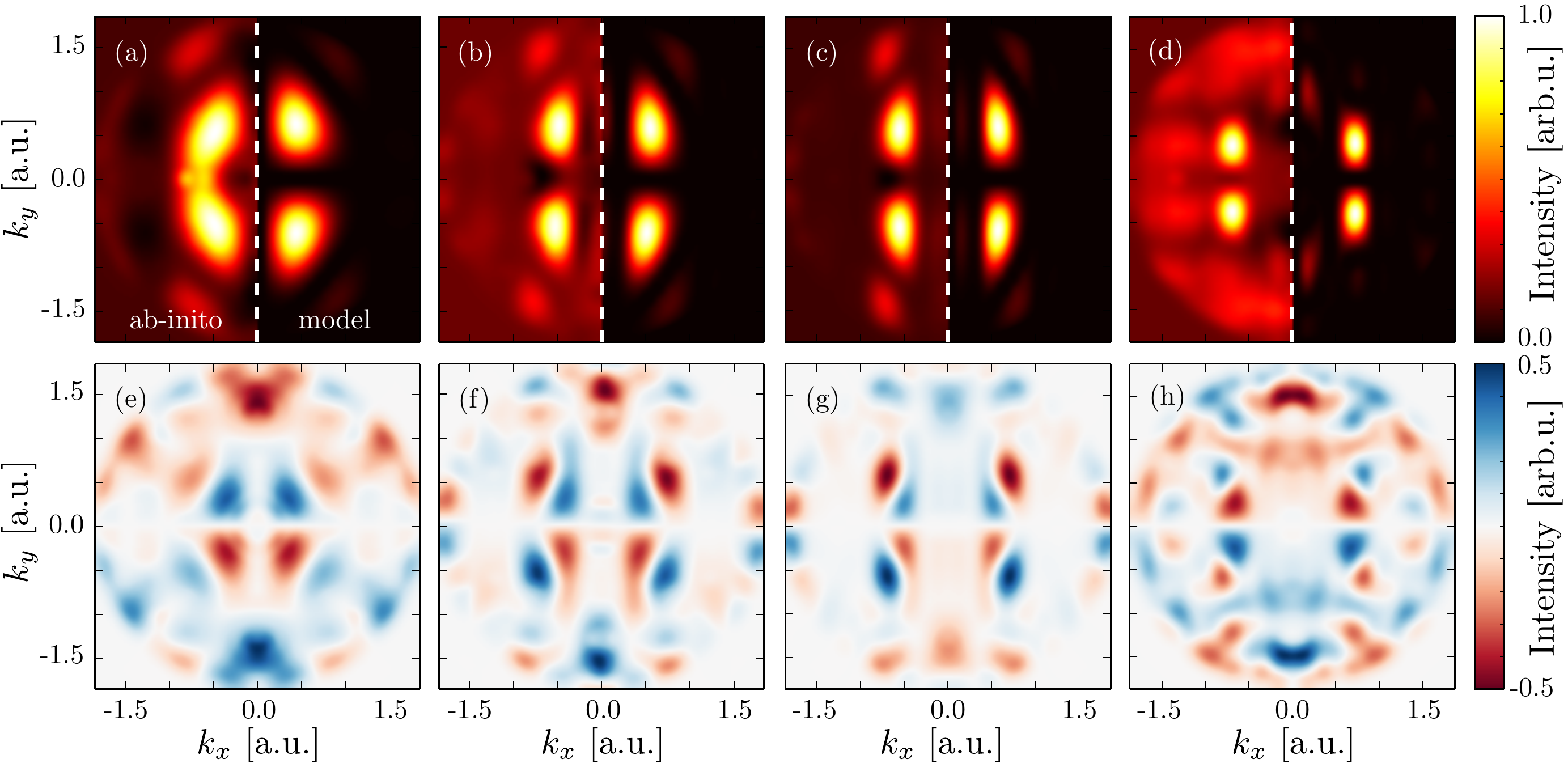}
  \caption{
  PADs and PECDs for a selection of organic molecules. From left to right in each column we report 
  results for a
  different molecule: naphtalene (a,e), anthracene (b,f), tetracene (c,g) , and PTCDA (d,h). 
  In the top row PADs are obtained using an unpolarized field, see Eq.\ref{eq:puk}, with 
  $\omega=54.4$\,eV, 
  $\tpulse=9$\,fs, and $I=10^8$\,W/cm$^2$. 
  For all the molecules we oriented the axis such that the $x$ ($y$) axis is along the 
  longest (shortest) 
  molecular axis while the $z$ axis is perpendicular to the plane of the molecule.
  Each panel is split in two: in the left part we plot the PAD obtained by the ab-initio simulation 
  using the mask method evaluated on the energy shell corresponding to electrons associated with 
  the HOMO (see text), 
  on the right we plot the Fourier transform of the HOMO orbital evaluated on the same energy shell.
  In the bottom row we show PECD maps resulting by subtracting the PADs obtained with left and right 
  circularly polarized 
  pulses. The laser has the same parameters as the one used for the top row but circularly 
  polarized on the $z$-$y$ plane. 
  }
  \label{fig:pah_pecd}
\end{figure}

It must be noted that the aforementioned relation is not universal and is supposed to be valid 
only for a limited set of molecules and orbitals ~\cite{Puschnig:2009ho}. 
The class of planar organic molecules satisfy these conditions and is thus well suited to 
illustrate the concept. 
To this end, we calculated $P_U(\mathbf{k})$ ab-initio for a selection of organic molecules: 
naphtalene, anthracene, tetracene, and perylenetetracarboxylic dianhydride (PTCDA).
The results are reported in Fig.~\ref{fig:pah_pecd}~(a--d) where we used a laser pulse 
with $\omega=54.4$~eV, $\tpulse=9$~fs, and $I=10^8$~W/cm$^2$ and cut at the energy shells 
corresponding to the HOMO for each molecule, namely 
$E_H=46.2$, 47.2, 47.8, 46.6~eV, respectively.
All the photoelectron calculations were performed using MM.

In each panel we split the figure in two parts and directly compare $P_U(\mathbf{k})$ (on the left) 
to the Fourier transform of the HOMO orbital $|\tilde{\varphi}_{\rm HOMO}(\mathbf{k})|^2$ (on the right). 
By comparing right and left parts of 
each panel it is clear that Eq.~\ref{eq:puk} well describes each photoelectron distribution.
In a first approximation all these molecules exhibit a similar structure and differ mainly by the 
number of phenyl rings. For this reason it is not surprising that all the PAD look similar 
except for small features forming close to zero momentum. This can be indeed understood in terms of 
Fourier transform where the lobes positioned at $k_y=\pm 1$\,a.u. can be associated with a dominant 
pattern well localized in space -- the phenyl ring. This base pattern is repeated for an increasing 
number of times (and in different directions) as we increase the size of the molecule, and this 
contributes to create small features in the short wavelengths in reciprocal space (large extension 
in real space). 

Once established the validity of Eq.~\ref{eq:puk} it is a natural step to attempt a Fourier 
inversion and reconstruct the orbitals in real space from the photoemission data. 
This, however, is not possible since $P_U(\mathbf{k})$ provides only information on the square 
modulus of the Fourier transform and we lack information about its phase. 
There have been different attempts in the literature to address this problem and to a large extent 
they reduce to different levels of educated guessing. For instance, in Ref.~\cite{Puschnig:2009ho},  
the phase was arbitrarily chosen, while in Ref.~\cite{Luftner:2014ib} it was selected with a 
self-consistent procedure and in Ref.~\cite{Wiessner:2014kq} it was identified by correlating 
group-symmetry arguments with photoelectron circular dichroism (PECD).

In what follows we focus on the last approach. 
To this end, we calculate the PECD maps subtracting the PAD obtained with left ($\sigma_+$) and 
right ($\sigma_-$) circularly polarized laser pulses 
$P_{CD}(\mathbf{k})= P_{+}(\mathbf{k})-P_{-}(\mathbf{k})$.
For the calculations we choose a field polarized on the $y$-$z$ axis provided that all the molecules 
are 
oriented with the 
longer (shorter) molecular axis on along $x$ ($y$) and that $z$ is perpendicular to the molecular 
plane.
The results are shown in Fig.~\ref{fig:pah_pecd} (e--h).
Given that none of the molecules in our set have a specific handedness one would expect to see 
zero dichroism, but clearly 
the PECD maps in the figure are not. The apparent discrepancy comes from the fact that it is our 
observation setup that 
has a defined handedness and therefore we can observe a dichroic effect even on molecules without 
a specifc handedness~\cite{Schonhense:1990eb}.
Furthermore, the results for PTCDA in Fig.~\ref{fig:pah_pecd}~(d) and (h) are in good agreement 
with the one measured in 
Ref.~\cite{Wiessner:2014kq}, especially considering that the experiment was carried out with 
molecules deposited on a metallic surface while our calculations are in the vacuum.

We conclude by observing that the final wave approximation we made to derive Eq.~\ref{eq:pk}, is 
inconsistent with a non-zero PECD. In fact, for circularly polarized light, Eq.~\ref{eq:pk} becomes 
\begin{equation}
  P_\pm(\mathbf{k})\propto |\langle \mathbf{k}| A \hat{p}_y \pm i A \hat{p}_z| \varphi_i\rangle |^2
  = P_y(\mathbf{k})+P_z(\mathbf{k})\pm i (D_y^*D_z - D_yD_z^*)\,
\end{equation}
where $D_{y,z}=\langle \mathbf{k}| A \hat{p}_{y,z}  |\varphi_i\rangle$ is the dipole matrix element.
The resulting PECD is $P_{CD}(\mathbf{k}) = 4 |D_y||D_z|\sin( \angle D_z-\angle D_y)$ and depends 
on the phase difference between the two dipole matrix elements of $y$ and $z$.
Approximating the final wavefunction as a plane wave implies that the matrix element 
$D_{y,z}=Ak_{y,z}\tilde{\varphi}(\mathbf{k})$ has a phase which is independent of the direction 
$y$,$z$ and thus $P_{CD}(\mathbf{k})=0$.
In order to have a non-zero PECD one has to go beyond the 
single plane wave final state approximation. Going beyond this approximation may also disclose 
information on the phase intrinsically encoded in the PECD and possibly allow Fourier inversion 
without the need of additional information on the group symmetry of the molecule. 
However, interesting, further investigation along this line is beyond the scope of the current paper. 
\section{Discussion}
Among the presented approaches to calculate photoelectron spectra with TDDFT, the sampling point 
method appears to be the 
most straightforward one to implement without effecting much the computational time.
However, the errors of the SPM may become severe since the 
quiver motion of the electron in the laser field is described only approximately at the 
position of the sampling point. This might lead to inaccurate photoelectron 
spectra in the strong-field regime. To avoid this, 
one is forced to choose the size of the box such that electron flow and laser field 
do not overlap at the sampling point.
For example, an electron with kinetic energy of 125\,eV travels a distance of 
$\approx\,125\,$a.u.\ within one femtosecond. Therefore, one would need to choose 
box sizes of hundreds of Bohr in order to obtain reliable spectra of systems exposed to 
laser pulses consisting of only a few cycles. This basically rules out 
a computationally efficient usage of the SPM in three dimensions for strong fields.
With the PA-SPM the situation improves, but still the results can be inaccurate 
if the photoelectrons have a non-negligible transversal momentum component at the sampling point.

The mask method, in principle, describes the time evolution of independent particles 
in the analyzed region (for short-range potentials) exactly as it projects the orbitals 
onto Volkov states which include the correct phase.
This avoids the problem of overlapping electron flow and laser field and 
allows to choose box sizes which only have to accomodate the quiver motion.
Moreover, MM also allows electrons to come back from region $B$. 
However, Fourier transforms of the single-particle wavefunctions 
in the absorbing region are involved. Thus, periodic boundaries are automatically imposed 
and the resolution in kinetic-energy space depends on grid spacing and width of the 
absorbing zone. Additionally, the implementation of absorbing boundaries has to be done 
through a mask function, although it also can be cast in terms of an additional 
imaginary potential in the Schr\"odinger equation. 

The surface flux method can be seen as a combination of SPM and MM rectifying the 
disadvantages of both methods. In contrast to the SPM, \tsurff\ can be 
properly derived within a TDDFT formalism. 
Like in the MM, box sizes can be reduced to 
approximately the range of the electron quiver amplitude as the quiver motion 
can be described reliably at the position 
where the spectrum is evaluated (i.e., the surface). 
On the other hand, the grid in momentum space can be chosen 
arbitrarily up to an energy range of 
$E_\mathrm{max}\lesssim \min\{\pi/\Delta t, (\pi/\Delta x)^2/2\}$,
where $\Delta t$ are the time step and $\Delta x$ the grid spacing 
used in the numerical computation. The surface is transparent 
which means that it allows electrons from region $A$ entering region $B$ and vice 
versa and that it can be combined with any kind of boundary condition.
The calculation of the flux includes the plane wave factor $e^{\imath\veck\vecr}$ and 
spatial derivatives of the orbitals that need to be evaluated on a closed surface. Therefore, 
interpolation may be required and the implementation of the 
\tsurff\ method is more involved than in the SPM. 
However, it can be fully parallelized in grid points and orbitals.

\section{Conclusion}
In this paper, 
we have reviewed the theoretical methods for the calculation of photoelectron spectra within 
TDDFT: the sampling point method~(SPM), the time-dependent surface flux 
method~\tsurff, and the 
mask method~(MM). 
While SPM and MM were already established in the framework of TDDFT, \tsurff\ is 
new and has been so far employed in conjunction with other theory levels.
We exported \tsurff\ to TDDFT using a novel derivation in terms of the flux of the current density 
operator and discussed how the expansion in spherical harmonics can be crucial for an efficient 
implementation in real space codes.  

We presented a direct comparison of all the three methods. Our benchmark was the simulation of the 
characteristic photoelectron angular distribution (PAD) 
of ATI peaks in an hydrogen atom. In our test, \tsurff\ 
emerged as the best method combining the flexibility and light computational cost of the SPM with the 
accuracy of the MM in excellent agreement with previously published results.

With \tsurff\ we investigated electron emission from the C$_{60}$ molecule exposed to 
an intense IR laser field. 
To the best of our knowledge this is the first time that a  
TDDFT atomistic simulation of strong-field ionization of such a large molecule is 
presented in the literature. 
The PES can be separated into 
direct and rescattered parts with a smooth plateau up to 
a cut-off at around $E_\mathrm{cut}^\mathrm{SFA}=10U_p$. 
The calculated angular distributions can be well explained with the quantitative 
rescattering theory and exhibit enhanced sidewards scattering which is completely missing in the same 
simulation with a jellium model.
This underlines the relevance of a theoretical description of strong-field phenomena
at the atomistic level.

Finally, we discussed the problem of orbital reconstruction from photoemission data. 
To this end, we performed simulations with both linearly and circularly polarized pulses on different 
planar organic molecules. We illustrated how the PAD from unpolarized fields is strongly connected 
with the Fourier transform of the molecular orbital from which the electrons originate and discussed 
the problem of Fourier inversion to recover the orbital from photoelectron data. 
We also performed photoelectron circular dichroism (PECD) simulations that are in good agreement 
with published data. Furthermore, we showed how non-zero PECD is in direct contrast with perturbation 
theory models in which the final state is approximated with a single plane wave and discussed how, 
going beyond this approximation, may constitute a possible venue to systematically recover the 
phase for orbital reconstruction.

\section*{Acknowledgments}
We acknowledge financial support from the European Research Council (ERC-2010-AdG-267374), Spanish
grant (FIS2013-46159-C3-1-P) and  Grupos Consolidados (IT578-13).
This work was partly supported by the European Union's Horizon 2020 
research and innovation program under grant agreement no. 676580 with the Novel 
Materials Discovery (NOMAD) laboratory, a European Center of Excellence,
H2020-NMP-2014 project MOSTOPHOS (GA no. 646259), and the COST Action MP1306 (EUSpec).
Finally, we acknowledge B.~Frusteri for his valuable help in testing the code. 

\section*{Appendix A}\label{sec:app_a}
The main complication for a computationally efficient implementation 
of Eq.~\ref{eq:flux} in three dimensions is the plane wave factor $e^{\imath\veck\vecr}$ in 
the Volkov state Eq.~\ref{eq:volkovstate} as the number of surface points and the 
$\veck$-grid become large. Therefore, we integrate on a sphere 
with $d^2\sigma = r_S^2\,d\Omega_r$ and expand the plane wave factor in spherical 
harmonics: 

\begin{eqnarray}\label{eq:sphflux}
  b_i(\veck) &= -\frac{r_S^2}{2\imath} \int dt\int d\Omega_r\,
  \chi_\veck^*\left[\nabla_i+\imath\veck-2\imath\frac{\veca}{c}\right]\varphi_i^A\cdot\vece_r \nonumber\\
  & = -\frac{4\pi r_S^2}{2\imath (2\pi)^{3/2}}
  \sum\limits_{lm}(-\imath)^l j_l(kr)Y_{lm}(\Omega_k)
  \int dt\,e^{\imath\Phi(\veck,t)}\times \\
  &\qquad\left\{\left(\imath\veck-2\imath\frac{\veca}{c}\right)
  \underbrace{\int d\Omega_r\,Y_{lm}^*(\Omega_r)\,\vece_r\,\varphi_i^A}_{\mathbf{S}_{lm}^{(1)}(t)} +
  \underbrace{\int d\Omega_r\,Y_{lm}^*(\Omega_r)\,\partial_r\varphi_i^A}_{S_{lm}^{(2)}(t)}
  \right\}\:.\nonumber
\end{eqnarray}
The expansion has the strong advantage that it decouples the surface integrals 
$\mathbf{S}_{lm}^{(1)}(t)$ and $S_{lm}^{(2)}(t)$ from the $\veck$-grid. The integration 
can be performed efficiently by using the Gaussian quadrature on interpolated Gaussian 
nodes up to a cutoff angular momentum~$L_\mathrm{max}$. 
For a given~$L_\mathrm{max}$, the number of integration nodes is independent of the 
size of the computational box.

\section*{Appendix B}\label{sec:app_b}

The connection between \tsurff\ and PA-SPM is to skip the surface integral in Eq.~\ref{eq:flux} 
and to consider 
only a single point, namely the one 
pointing in $\veck$-direction. In three dimensions, we proceed by
replacing 
$\mathbf{J}_\veck^{(i)}\rightarrow\mathbf{J}_\veck^{(i)}\cdot \delta^{(2)}(\Omega_k-\Omega_r)$
and obtain:
\begin{eqnarray}
  b_i^\mathrm{alt}(\veck) &= -\int\limits_0^T dt\,\mathbf{J}_\veck^{(i)}(r,\Omega_k;t)\cdot\vece_k 
  \nonumber\\
  &=-\frac{e^{-\imath kr}}{2\imath (2\pi)^{3/2}}\int dt\,e^{\imath\Phi(\veck;t)}
  \left[\nabla_i+\imath\veck-2\imath\frac{\veca}{c}\right]\varphi_i(r,\Omega_k;t)\cdot\vece_k
  \label{eq:spa3d}
\end{eqnarray}
Assuming $\partial_r\tilde{\xi}(r,\Omega_k;\omega)=\sqrt{2\omega}\imath\,
\tilde{\xi}(r,\Omega_k;\omega)$ and 
identifying $k=\sqrt{2\omega}$, we recover (except for a normalization factor and the term
accounting for the vector potential) the result 
from the PA-SPM.

\section*{References}
% \bibliography{library}

\begin{thebibliography}{10}
\expandafter\ifx\csname url\endcsname\relax
  \def\url#1{{\tt #1}}\fi
\expandafter\ifx\csname urlprefix\endcsname\relax\def\urlprefix{URL }\fi
\providecommand{\eprint}[2][]{\url{#2}}

\bibitem{Tao12}
Tao L and Scrinzi A 2012 {\em New Journal of Physics\/} {\bf 14} 013021

\bibitem{Kel65}
Keldysh L~V 1965 {\em Soviet Physics, JETP\/} {\bf 20} 1307

\bibitem{Man91}
Mainfray G and Manus G 1991 {\em Reports on Progress in Physics\/} {\bf 54}
  1333--1372

\bibitem{Ago79}
Agostini P, Fabre F, Mainfray G, Petite G and Rahman N 1979 {\em Physical
  Review Letters\/} {\bf 42} 1127--1130

\bibitem{Bra00}
Brabec T and Krausz F 2000 {\em Reviews of Modern Physics\/} {\bf 72} 545

\bibitem{Cor93}
Corkum P~B 1993 {\em Physical Review Letters\/} {\bf 71} 1994--1997

\bibitem{Kul93}
Kulander K~C, Schafer K~J and Krause K~L 1993 {Super-Intense Laser-Atom
  Physics} {\em NATO Advanced Study Institute: Series B: Physics\/} ed Piraux
  B, L'Huillier A and Rzazewski K (Plenum, New York) p~95

\bibitem{Lew95}
Lewenstein M, Kulander K~C, Schafer K and Bucksbaum P 1995 {\em Physical Review
  A\/} {\bf 51} 1495--1507

\bibitem{Fai73}
Faisal F~H~M 1973 {\em Journal of Physics B: Atomic and Molecular Physics\/}
  {\bf 6} L89--L92

\bibitem{Rei80}
Reiss H~R 1980 {\em Physical Review A\/} {\bf 22} 1786--1813

\bibitem{Kra09}
Krausz F and Ivanov M 2009 {\em Reviews of Modern Physics\/} {\bf 81} 163

\bibitem{Mec08}
Meckel M, Comtois D, Zeidler D, Staudte A, Pavicic D, Bandulet H~C, Pepin H,
  Kieffer J~C, Dorner R, Villeneuve D~M and Corkum P~B 2008 {\em Science\/}
  {\bf 320} 1478--1482

\bibitem{Bla12}
Blaga C~I, Xu J, DiChiara A~D, Sistrunk E, Zhang K, Agostini P, Miller T~A,
  DiMauro L~F and Lin C~D 2012 {\em Nature\/} {\bf 483} 194--197

\bibitem{Huismans2011}
Huismans Y, Rouzee A, Gijsbertsen A, Jungmann J~H, Smolkowska A~S, Logman
  P~S~W~M, L{\'{e}}pine F, Cauchy C, Zamith S, Marchenko T, Bakker J~M, Berden
  G, Redlich B, van~der Meer A~F~G, Muller H~G, Vermin W, Schafer K~J, Spanner
  M, Ivanov M~Y, Smirnova O, Bauer D, Popruzhenko S~V and Vrakking M~J~J 2011
  {\em Science\/} {\bf 331} 61--65

\bibitem{Spa04}
Spanner M, Smirnova O, Corkum P~B and Ivanov M~Y 2004 {\em Journal of Physics
  B: Atomic, Molecular and Optical Physics\/} {\bf 37} L243--L250

\bibitem{Rac11}
R{\'{a}}cz P, Irvine S~E, Lenner M, Mitrofanov A, Baltu{\v{s}}ka A, Elezzabi
  A~Y and Dombi P 2011 {\em Applied Physics Letters\/} {\bf 98} 111116

\bibitem{Dom13a}
Dombi P, H{\"{o}}rl A, R{\'{a}}cz P, M{\'{a}}rton I, Tr{\"{u}}gler A, Krenn J~R
  and Hohenester U 2013 {\em Nano Letters\/} {\bf 13} 674

\bibitem{Fen07}
Fennel T, D{\"{o}}ppner T, Passig J, Schaal C, Tiggesb{\"{a}}umker J and
  Meiwes-Broer K~H 2007 {\em Physical Review Letters\/} {\bf 98} 143401

\bibitem{Zhe11}
Zherebtsov S, Fennel T, Plenge J, Antonsson E, Znakovskaya I, Wirth A,
  Herrwerth O, S{\"{u}}{\ss}mann F, Peltz C, Ahmad I, Trushin S~A, Pervak V,
  Karsch S, Vrakking M~J~J, Langer B, Graf C, Stockman M~I, Krausz F,
  R{\"{u}}hl E and Kling M~F 2011 {\em Nature Physics\/} {\bf 7} 656--662

\bibitem{Pig13a}
Piglosiewicz B, Schmidt S, Park D~J, Vogelsang J, Gro{\ss} P, Manzoni C,
  Farinello P, Cerullo G and Lienau C 2013 {\em Nature Photonics\/} {\bf 8}
  37--42

\bibitem{Kru11}
Kr{\"{u}}ger M, Schenk M and Hommelhoff P 2011 {\em Nature\/} {\bf 475} 78--81

\bibitem{Her12a}
Herink G, Solli D~R, Gulde M and Ropers C 2012 {\em Nature\/} {\bf 483} 190--3

\bibitem{Bio13}
Bionta M~R, Chalopin B, Champeaux J~P, Faure S, Masseboeuf A, Moretto-Capelle P
  and Chatel B 2013 {\em Journal of Modern Optics\/}  1--6

\bibitem{Bac01}
Bachau H, Cormier E, Decleva P, Hansen J~E and Mart{\'{i}}n F 2001 {\em Reports
  on Progress in Physics\/} {\bf 64} 1815--1943

\bibitem{Cat12}
Catoire F and Bachau H 2012 {\em Physical Review A\/} {\bf 85} 023422

\bibitem{Che98}
Chelkowski S, Foisy C and Bandrauk A~D 1998 {\em Physical Review A\/} {\bf 57}
  1176--1185

\bibitem{Ton07}
Tong X~M, Hino K, Toshima N and Burgd{\"{o}}rfer J 2007 {\em Journal of
  Physics: Conference Series\/} {\bf 88} 012047

\bibitem{He15}
He P~L, Takemoto N and He F 2015 {\em Physical Review A\/} {\bf 91} 063413

\bibitem{Che06}
Chen Z, Morishita T, Le A~T, Wickenhauser M, Tong X~M and Lin C~D 2006 {\em
  Physical Review A\/} {\bf 74} 053405

\bibitem{Awa08}
Awasthi M, Vanne Y~V, Saenz A, Castro A and Decleva P 2008 {\em Physical Review
  A\/} {\bf 77} 1--17

\bibitem{Pet10}
Petretti S, Vanne Y~V, Saenz A, Castro A and Decleva P 2010 {\em Physical
  Review Letters\/} {\bf 104} 2--5

\bibitem{Mad97}
Madsen L~B and Plummer M 1998 {\em Journal of Physics B: Atomic, Molecular and
  Optical Physics\/} {\bf 31} 87--104

\bibitem{Chu04}
Chu S~I and Telnov D~A 2004 {\em Physics Reports\/} {\bf 390} 1--131

\bibitem{MB00}
Muth-B{\"{o}}hm J, Becker A and Faisal F~H~M 2000 {\em Physical Review
  Letters\/} {\bf 85} 2280--2283

\bibitem{Dre14}
Dreissigacker I and Lein M 2014 {\em Physical Review A\/} {\bf 89} 053406

\bibitem{Yud01}
Yudin G~L and Ivanov M~Y 2001 {\em Physical Review A\/} {\bf 63} 1--14

\bibitem{Ton02}
Tong X~M, Zhao Z~X and Lin C~D 2002 {\em Physical Review A\/} {\bf 66} 033402

\bibitem{Dim04}
Dimitriou K~I, Arb{\'{o}} D~G, Yoshida S, Persson E and Burgd{\"{o}}rfer J 2004
  {\em Physical Review A\/} {\bf 70} 1--4

\bibitem{Eck08}
Eckle P, Pfeiffer a~N, Cirelli C, Staudte A, D{\"{o}}rner R, Muller H~G,
  B{\"{u}}ttiker M and Keller U 2008 {\em Science (New York, N.Y.)\/} {\bf 322}
  1525--9

\bibitem{ADK86}
Ammosov M~V, Delone N~B and Krainov V 1986 {\em Sov Phys JETP\/} {\bf 64} 1191

\bibitem{Pus09}
Puschnig P, Berkebile S, Fleming A~J, Koller G, Emtsev K, Seyller T, Riley J~D,
  Ambrosch-Draxl C, Netzer F~P and Ramsey M~G 2009 {\em Science (New York,
  N.Y.)\/} {\bf 326} 702--6

\bibitem{Gro78}
Grobman W~D 1978 {\em Physical Review B\/} {\bf 17} 4573--4585

\bibitem{Tof12}
Toffoli D and Decleva P 2012 {\em The Journal of chemical physics\/} {\bf 137}
  134103

\bibitem{Sei02}
Seideman T 2002 {\em Annual review of physical chemistry\/} {\bf 53} 41--65

\bibitem{Fai87}
Faisal F~H~M 1987 {\em {Theory of multiphoton processes}\/} (Plenum Press, New
  York)

\bibitem{RG84}
Runge E and Gross E~K~U 1984 {\em Physical Review Letters\/} {\bf 52} 997--1000

\bibitem{Fun12}
Marques M~A~L, Maitra N~T, Nogueira F~M~S, Gross E~K~U and Rubio A 2012 {\em
  {Fundamentals of Time-Dependent Density Functional Theory}\/}
  (Springer-Verlag, Berlin) ISBN 9783642235177

\bibitem{For92}
Foresman J~B, Head-Gordon M, Pople J~A and Frisch M~J 1992 {\em The Journal of
  Physical Chemistry\/} {\bf 96} 135--149

\bibitem{Zan04}
Zanghellini J, Kitzler M, Brabec T and Scrinzi A 2004 {\em Journal of Physics
  B: Atomic, Molecular and Optical Physics\/} {\bf 37} 763--773

\bibitem{Dauth2011}
Dauth M, K{\"{o}}rzd{\"{o}}rfer T, K{\"{u}}mmel S, Ziroff J, Wiessner M,
  Sch{\"{o}}ll A, Reinert F, Arita M and Shimada K 2011 {\em Physical Review
  Letters\/} {\bf 107} 193002

\bibitem{Larsen:2015kc}
Larsen A~H, De~Giovannini U and Rubio A 2015 {Dynamical Processes in Open
  Quantum Systems from a TDDFT Perspective: Resonances and Electron
  Photoemission} {\em Density-Functional Methods for Excited States\/} (Cham:
  Springer International Publishing) pp 219--271

\bibitem{Poh00}
Pohl A, Reinhard P~G and Suraud E 2000 {\em Physical Review Letters\/} {\bf 84}

\bibitem{Di12b}
Dinh P~M, Romaniello P, Reinhard P~G and Suraud E 2013 {\em Physical Review
  A\/} {\bf 87} 032514

\bibitem{deGio12}
de~Giovannini U, Varsano D, Marques M~A~L, Appel H, Gross E~K~U and Rubio A
  2012 {\em Physical Review A\/} {\bf 85} 62515

\bibitem{Wop14a}
Wopperer P, Dinh P~M, Reinhard P~G and Suraud E 2015 {\em Physics Reports\/}
  {\bf 562} 1--68

\bibitem{Wac12a}
Wachter G, Lemell C and Burgd{\"{o}}rfer J 2012 {\em Journal of Physics:
  Conference Series\/} {\bf 399} 012010

\bibitem{DeGio13}
De~Giovannini U, Brunetto G, Castro A, Walkenhorst J and Rubio A 2013 {\em
  Chemphyschem\/} {\bf 14} 1363--76

\bibitem{CrU14b}
Crawford-Uranga A, {De Giovannini} U, Mowbray D~J, Kurth S and Rubio A 2014
  {\em Journal of Physics B: Atomic, Molecular and Optical Physics\/} {\bf 47}
  124018

\bibitem{CrU14a}
Crawford-Uranga A, {De Giovannini} U, R{\"{a}}s{\"{a}}nen E, Oliveira M~J~T,
  Mowbray D~J, Nikolopoulos G~M, Karamatskos E~T, Markellos D, Lambropoulos P,
  Kurth S and Rubio A 2014 {\em Physical Review A\/} {\bf 90} 033412

\bibitem{Scr12}
Scrinzi A 2012 {\em New Journal of Physics\/} {\bf 14} 085008

\bibitem{Cai05}
Caillat J, Zanghellini J, Kitzler M, Koch O, Kreuzer W and Scrinzi A 2005 {\em
  Physical Review A\/} {\bf 71} 012712

\bibitem{Morales:2016va}
Morales F, Bredtmann T and Patchkovskii S 2016  (\textit{Preprint}
  \eprint{arxiv:1606.04566})

\bibitem{deGio14}
De~Giovannini U, Larsen A~H and Rubio A 2015 {\em The European Physical Journal
  B\/} {\bf 88} 56

\bibitem{PW92}
Perdew J~P and Wang Y 1992 {\em Physical Review B\/} {\bf 45} 13244--13249

\bibitem{Leg02}
Legrand C, Suraud E and Reinhard P~G 2002 {\em Journal of Physics B: Atomic,
  Molecular and Optical Physics\/} {\bf 35} 1115

\bibitem{And15}
Andrade X, Strubbe D, {De Giovannini} U, Larsen A~H, Oliveira M~J~T,
  Alberdi-Rodriguez J, Varas A, Theophilou I, Helbig N, Verstraete M~J, Stella
  L, Nogueira F, Aspuru-Guzik A, Castro A, Marques M~A~L and Rubio A 2015 {\em
  Phys. Chem. Chem. Phys.\/} {\bf 17} 31371--31396

\bibitem{Cas06}
Castro A, Appel H, Oliveira M, Rozzi C~A, Andrade X, Lorenzen F, Marques M~A~L,
  Gross E~K~U and Rubio A 2006 {\em physica status solidi (b)\/} {\bf 243}
  2465--2488

\bibitem{Arb06}
Arb{\'{o}} D~G, Yoshida S, Persson E, Dimitriou K~I and Burgd{\"{o}}rfer J 2006
  {\em Physical Review Letters\/} {\bf 96} 143003

\bibitem{Zhou11}
Zhou Z and Chu S~I 2011 {\em Physical Review A - Atomic, Molecular, and Optical
  Physics\/} {\bf 83} 1--9

\bibitem{Gao16}
Gao C~Z, Dinh P~M, Kl{\"{u}}pfel P, Meier C, Reinhard P~G and Suraud E 2016
  {\em Physical Review A\/} {\bf 93} 022506

\bibitem{Lin10}
Lin C~D, Le A~T, Chen Z, Morishita T and Lucchese R 2010 {\em Journal of
  Physics B: Atomic, Molecular and Optical Physics\/} {\bf 43} 122001

\bibitem{Xu10}
Xu J, Chen Z, Le A~T and Lin C~D 2010 {\em Physical Review A\/} {\bf 82} 1--13

\bibitem{TM91}
Troullier N and Martins J~L 1991 {\em Physical Review B\/} {\bf 43} 1993--2006

\bibitem{Che09}
Chen Z, Le A~T, Morishita T and Lin C~D 2009 {\em Physical Review A\/} {\bf 79}
  1--18

\bibitem{Puschnig:2009ho}
Puschnig P, Berkebile S, Fleming A~J, Koller G, Emtsev K, Seyller T, Riley J~D,
  Ambrosch-Draxl C, Netzer F~P and Ramsey M~G 2009 {\em Science\/} {\bf 326}
  702--706

\bibitem{Luftner:2014ib}
L{\"u}ftner D, Ules T, Reinisch E~M, Koller G, Soubatch S, Tautz F~S, Ramsey
  M~G and Puschnig P 2014 {\em Proceedings Of The National Academy Of Sciences
  Of The United States Of America\/} {\bf 111} 605--610

\bibitem{Wiessner:2014kq}
Wie{\ss}ner M, Hauschild D, Sauer C, Feyer V, Sch{\"o}ll A and Reinert F 2014
  {\em Nature Communications\/} {\bf 5}

\bibitem{Schonhense:1990eb}
Sch{\"o}nhense G 2007 {\em Physica Scripta\/} {\bf T31} 255--275

\end{thebibliography}
% \bibliographystyle{iopart-num}
\providecommand{\newblock}{}

\end{document}